\documentclass[12pt]{revtex4}
\usepackage{rotating}
\usepackage{graphicx}% Include figure files

\usepackage{amsfonts}
\usepackage{amssymb}
\usepackage{amsmath}
\usepackage{latexsym}

\def\beq{\begin{equation}}
\def\eeq{\end{equation}}

\def\bq{\begin{quote}}
\def\eq{\end{quote}}

\newcommand{\non}{\nonumber}
\newcommand{\be}{\begin{equation}}
\newcommand{\ee}{\end{equation}}
\newcommand{\bea}{\begin{eqnarray}}
\newcommand{\eea}{\end{eqnarray}}
\newcommand{\ba}{\begin{array}}
\newcommand{\ea}{\end{array}}
\newcommand{\al}{\alpha}

\newcommand{\la}{\lambda}

\newcommand{\de}{\delta}

\newcommand{\De}{\Delta}
\newcommand{\vphi}{\varphi}

\newcommand{\rar}{\rightarrow}

\begin{document}

\title{\Large Double Well Potential: Perturbation Theory,\\[10pt]
 Tunneling, WKB \\(beyond instantons)}

\author{\large Alexander V Turbiner\footnotemark \footnotetext{
\uppercase{E}-mail: turbiner@nucleares.unam.mx} }

\address{ Instituto de Ciencias Nucleares, UNAM, A.P. 70-543,
 Mexico D.F., 04510 Mexico}

\begin{flushright}

 July 25, 2009
\end{flushright}

%\pacs{math-ph/0506033}

\maketitle

\begin{center}
{\bf\large Abstract}
\end{center}
\small{
\begin{quote}
A simple approximate solution
for the quantum-mechanical quartic oscillator
$V= m^2 x^2+g x^4$ in the double-well regime $m^2<0$ at arbitrary $g \geq 0$ is presented. 
It is based on a combining of perturbation theory near true minima of the potential, 
semi-classical approximation at large distances and a description of tunneling under 
the barrier. It provides 9-10 significant digits in energies and gives for wavefunctions
the relative deviation in real $x$-space less than $\lesssim 10^{-3}$.
\end{quote}}

\vskip 2cm

\newpage

Needless to say that since the creation of quantum mechanics the one-dimensional quartic
anharmonic oscillator
\begin{equation}
\label{AHO}
    {\cal H}\ =\ - \frac{d^2}{dx^2}\ +\ m^2 x^2 + g x^4\ ,
\end{equation}
always attracted a lot of attention being among the most celebrated
problem of quantum mechanics. The interest to these
problems ranges from various branches of physics, from quantum
field theory to chemistry and biology. It is especially true for the case
when $m^2 <0$ and the potential has two minima. This problem is known
in literature as the {\it double-well potential}. It was studied in hundreds
papers, appeared practically in all books on quantum mechanics. A special emphasis
was made to a domain $m^2 \rar \infty$ - a domain where the barrier penetration 
is described by instantons (see e.g. \cite{Polyakov:1977, Coleman:1979, Vainshtein:1981})
that gives rise to the instanton physics.

The first detailed study of (1) carried out by Bender-Wu at 1969-1973
\cite{BW} revealed in this seemingly simple Hamiltonian the extremely rich
analytic structure which looks intrinsic for any non-trivial eigenvalue problem
of quantum mechanics and even though for quantum field theory.
In fact, one of the most important unwritten conclusions was that
in no way this problem can be solved exactly. The goal of the present
talk is to give an approximate solution valid, actually, for any $g >0$ and $m^2$.
The solution is given in a form of the fairly simple expression for the ground state
and the first excited state wavefunctions, which for any real $x$ and for
$g \geq 0$, real $m^2$ differs from the exact wavefunction for not
more than a small number $\de$,
\[
  |\frac{\Psi_{approximate}-\Psi_{exact}}{\Psi_{approximate}}\vert \leq \delta \ .
\]
In our case the $\de \approx 10^{-6}$. Evidently, it implies that
{\it any} quantity related with the first two eigenstates like expectation values
can be calculated with accuracy $\delta^2$.

There are three basic analytic approaches to study the spectra in
quantum mechanics: (i) perturbation theory, (ii) WKB method and
(iii) instanton calculus 
\footnote{It must be noted that, in fact, any one-dimensional Schroedinger equation 
can be solved numerically with any desirable accuracy. However, it is not true for 
multidimensional case.}. 
Each approach has its domain of
applicability and usually these domains do not overlap.
We attempt to combine (incorporate, unify) all three approaches into one
by making interpolation. The most convenient object to incorporate (i) and (ii) is the 
{\it logarithmic derivative of the wavefunction}. While the suitable object to incorporate 
the property (iii) is the wavefunction. A final form of the approximation depends on a few 
free parameters. Roughly speaking, their behavior as a function of $g, m^2$ is rather smooth 
and simple. They can be fixed variationally, although it is not very important: a  small variation 
of the parameters does not lead to dramatic loss of accuracy.

As a first step to approach the problem let us remind the Symanzik rescaling for eigenvalues 
and eigenfunctions
\[
  E(m^2, g) = g^{1/3} E\bigg(\frac{m^2}{g^{2/3}}, 2\bigg)\ ,\
  \Psi (x; m^2, g) = \Psi \bigg(x g^{1/6}; \frac{m^2}{g^{2/3}}, 2\bigg)\ .
\]
It manifests that the original problem (1) is in fact one-parametric.
The Hamiltonian (\ref{AHO}) can be rewritten in the form
\begin{equation}
\label{AHOa}
    {\cal H}\ =\ - \frac{d^2}{dx^2}\ +\ a x^2 + 2 x^4\ ,\quad x \in (-\infty, +\infty)\ ,
\end{equation}
where $a \equiv \frac{m^2}{g^{2/3}}$. This is the form of the Hamiltonian we are
going to study with the real parameter $a$ varying from negative to positive values.
The Schroedinger equation for (\ref{AHOa}) reads
\begin{equation}
\label{SchAHOa}
    - \frac{d^2 \Psi}{dx^2}\ +\ a x^2 \Psi \ +\ 2 x^4\Psi\ =
    \ E\ \Psi \quad
    ,\quad \int_{-\infty}^{+\infty} |\Psi|^2 dx < \infty \ .
\end{equation}
Eigenfunctions of (\ref{SchAHOa}) are sharply changing functions in
$x \in {\bf R}$ being characterized by a power-like behavior
at $|x| \rar 0$ and an exponentially-decaying one
at $|x| \rar \infty$. Following the oscillation (Sturm) theorem the
$n$th eigenfunction has $n$ simple (real) zeros. It seems natural
to introduce the representation for eigenfunctions as follows
\begin{equation}
\label{psi}
 \Psi (x)\ =\ p_n(x) e^{-\vphi (x)}\ ,
\end{equation}
where the phase $\vphi (x)$ is a slow-changing smooth function and $p_n(x)$ is a polynomial 
of $n$th degree with real coefficients which has $n$ real roots
\cite{Turbiner:1980}. Recently, it was obtained a remarkable result \cite{Eremenko:2008}: Any 
eigenfunction $\Psi (x)$ of (2) for any real $a$ is entire function and it has infinitely-many 
simple complex zeros all situated on imaginary axis symmetrically! It implies that the phase $\vphi (x)$ has infinitely-many logarithmic branch points in complex $x$-plane and has no singularities at real $x$.

After substitution of (\ref{psi}) into (\ref{SchAHOa}) we get the following equation
\begin{equation}
\label{RicAHO}
    y' - y^2 - \frac{p''_n - 2y p'_n}{p_n}\ =\ E - \ a x^2\ -\ 2 x^4\ ,
    \quad y\ =\ \vphi'\ =\ (\log \Psi (x))'\ ,\ y(0)=0\ .
\end{equation}
In order to define the problem (\ref{RicAHO}) we impose two conditions that $y$ has no simple poles 
at real $x$ (i) and it grows at $|x| \rar \infty$ not faster than polynomial (ii). The condition (i) 
implies that the coefficients of the polynomial $p_n(x)$ are those that the residues in the simple 
poles in the third term in l.h.s. of (\ref{RicAHO}) vanish. The condition (ii) assures squire-integrability 
of the wavefunction (\ref{psi}). It is evident that the polynomial $p_n(x)$ has parity $p=(-)^n$, hence, 
can be written as
$p_n(x)=x^p P_{[\frac{n}{2}]} (x^2)$, and
$y(x)$ is odd, $y(x)=y(-x)$ (for discussion, see \cite{Turbiner:1984}).

From the analysis of (\ref{RicAHO}) it is easy to find asymptotic behavior of the phase 
\footnote{The phase is defined up to additive constant which is fixed following a normalization 
of the wavefunction. We will omit it.},
\begin{equation}
\label{y-infty}
  \vphi\ =\ \frac{2^{1/2}}{3} x^2 |x| + \frac{a}{2^{3/2}} |x| +
  (n+1) \log{|x|} + \frac{8 E + a^2}{2^{9/2}} \frac{1}{|x|} +
  \bigg(\frac{a}{2}-4A\bigg) \frac{1}{x^2}+
  \ldots \quad \mbox{at}\ |x| \rar \infty \ ,
\end{equation}
where $A$ is sum of squared of nodes, while
\begin{equation}
\label{y-zero}
  \vphi \ =\ \frac{E}{2}x^2 + \frac{E^2-a}{12} x^4 +
  \frac{2E(E^2-a)-6}{90}x^6 + \ldots \quad \mbox{at}\ |x| \rar 0
  \ ,
\end{equation}
(see \cite{Turbiner:2005}). It is important to note that the first two terms in (\ref{y-infty}) 
are defined by the equation (\ref{RicAHO}) with omitted $y'$ term, which is actually the Hamilton-Jacobi 
equation. Therefore, these two terms coincide with first two terms of the asymptotics of the classical 
action at $|x| \rar \infty$. The third term in (\ref{y-infty}) is also reproduced in the expansion of 
the classical action but with a wrong coefficient. The correct coefficient can be obtained if the first 
correction to the classical action is taken into account (quadratic fluctuations). The first three terms 
in (\ref{y-infty}) grow when $|x|$ tends to $\infty$. In fact, they characterize a singularity at $|x|=\infty$.
These terms do not depend on energy $E$ and are found explicitly. For the single well potential $a \geq 0$ 
the expansion (\ref{y-zero}) is nothing but the (divergent) perturbation theory series near the minimum of 
the potential. It is an expansion around the true vacuum.

Now let us construct a simplest function for phase (\ref{psi}) which interpolates small and large distance expansions, reproducing
{\it exactly} the first three (growing) terms in (\ref{y-infty}).
It has the form,
\begin{equation}
\label{phase_int}
  \vphi_{int}\ =\
  \frac{A + (D^2+3a)  x^2 + 4 x^4}{6 (D^2 + 2 x^2)^{1/2}}
  - \frac{n+1}{2} \log {(D^2 + 2 x^2)} \ .
\end{equation}
%%%  new A is equal to 6 multiply by old A.
The corresponding wavefunction (\ref{psi}) for $n \equiv (2k+p)$ excited state (for $k=0,1,\ldots$ and 
parity $p=0,1$)  is equal to
\begin{equation}
\label{psi0}
 \psi_{0}^{(k,p)}\ =\ \frac{x^p P_k(x^2)}{(D^2 + 2 x^2)^{k+\frac{p+1}{2}}}
   \exp\left\{
   -\frac{A + (D^2+3a)  x^2 + 4 x^4}{6 (D^2 + 2 x^2)^{1/2}}
       \right\}\ ,
\end{equation}
(cf. \cite{Turbiner:2005} at $k=0$), where $A,D$ are free parameters and $P_k$ is a polynomial with real 
coefficients having positive roots only.
There are two ways to find the polynomial $P_k$: either imposing the orthogonality conditions to the 
functions with smaller quantum numbers $k=0,1,\ldots (k-1)$, or requiring the absence of simple poles 
in the third term in l.h.s. of (\ref{RicAHO}). Surprisingly, in concrete calculations these two conditions 
lead to polynomials whose coefficients coincide with high accuracy. Each function $\psi_{0}^{(k,p)}$ is 
characterized by two free parameters $A,D$, which can be fixed if the function (\ref{psi0}) is taken as 
variational trial function. If $k=0$ in (\ref{psi0}) both ground ($p=0$) and the first excited ($p=1$) 
states occur, respectively. Concrete calculations of the variational energies of these states with 
(\ref{psi0}) taken as a trial function lead to unprecedented accuracy \cite{Turbiner:2005} (see below). 
Furthermore, if in the exponential in (\ref{psi0}) in the term $4 x^4$ (this term governs the asymptotic 
behavior of the wave function at large distances where the wave function is exponentially small) the 
factor 4 is replaced by a parameter, $W x^4$, minimization of such a trial function leads to a value 
of $W$ which is equal to 4 with accuracy $10^{-4}$!

In the case $a<0$ the vacuum at $x=0$ becomes the false vacuum and two classically degenerate vacua at 
$x_{\pm}= \pm a^{1/2}/2$ appear. The expansion (\ref{y-zero}) becomes the expansion around false minimum 
(maximum), it is not relevant physically. The expansion around one or another true vacuum should be 
considered instead, as a relevant one. This expansion can be easily derived and we skip it.

In this case a new physical phenomenon of the quantum mechanical tunneling (barrier penetration) 
occurs. There is a probability to meet the particle under the barrier, near $x \approx 0$. It decays 
exponentially when $a \rar -\infty$. This phenomenon is absent in (\ref{psi0}). A prescription how to 
describe tunneling it is given in the celebrated Landau-Lifschitz book \cite{LL-QM:1977}. It employs
a linear superpositions of the wavefunctions centered at different minima which has positive parity 
for the ground state and similar one of negative parity for the first excited state. From the viewpoint 
of construction of the interpolation it is similar to making an interpolation between the expansion at 
one minimum and another one, and at $|x| \rar \infty$. Finally, such an interpolating function is 
a linear superposition of two off-centered functions (\ref{psi0}), which can be written as
\begin{equation}
\label{psi0+}
 \psi_{0}^{(k,+)}\ =\ \frac{q^+_k(x^2)}{(D_+^2 + 2 x^2)^{k+\frac{1}{2}}}
   \cosh{\frac{\al_+ x}{(D_+^2 + 2 x^2)^{1/2}}} \exp\left\{
   -\frac{A_+ + (D_+^2+3a)  x^2 + 4 x^4}{6 (D_+^2 + 2 x^2)^{1/2}}
       \right\}\ ,
\end{equation}
for the states of positive parity and
\begin{equation}
\label{psi0-}
 \psi_{0}^{(k,-)}\ =\ \frac{q^-_k(x^2)}{(D_-^2 + 2 x^2)^{k+1}}
   \sinh{\frac{\al_- x}{(D_-^2 + 2 x^2)^{1/2}}} \exp\left\{
   -\frac{A_- + (D_-^2+3a)  x^2 + 4 x^4}{6 (D_-^2 + 2 x^2)^{1/2}}
       \right\}\ ,
\end{equation}
for the states of negative parity. Parameter $\al_+(\al_-)$ `measures' a 
displacement of the peaks of wave function from the origin, $x=0$. If $\al_+=0$ 
the function (\ref{psi0+}) becomes (\ref{psi0}). Here $(q^{\pm}_k)$ is a polynomial 
of degree $k$ with real coefficients having positive roots only. At fixed $k$ any 
function depends on three free parameters $\al,A,D$.

In order to proceed further I need to remind two important and poorly known results:
(i) a special form of perturbation theory in QM sometimes called `Logarithmic
Perturbation Theory' or 'Non-linearization Method', and (ii) a connection 
between variational calculation and perturbation theory.

({\bf i}) A special form of perturbation theory is a certain iterative
procedure developed for solving the Riccati equation (\ref{RicAHO})
instead of the Schroedinger equation. For finding the wave function
it is a multiplicative perturbation theory unlike a standard additive
Rayleigh-Schroedinger perturbation theory. Such a multiplicative
perturbation theory was developed for the first time by Price
\cite{Price} and then it was numerously rediscovered (for early
history and discussion see \cite{Turbiner:1984} and references therein).
In presentation we follow closely to \cite{Turbiner:1979}. For simplicity
we will consider the eigenstates for which the nodes are absent (ground state)
or nodal positions are known.

As a first step to develop the perturbation theory we make a choice
of some square-integrable function $\Psi_0$ and calculate its logarithmic
derivative
\begin{equation}
\label{y0}
 y_0 = (\log \Psi_0)' = \frac{\Psi_0'}{\Psi_0}\ .
\end{equation}
It is clear that $\Psi_0$ is the exact eigenfunction of the
Schroedinger operator with a potential
\begin{equation}
\label{V0}
  V_0 = \frac{\Psi_0''}{\Psi_0} = y_0^2 - y_0'\ ,
\end{equation}
where without a loss of generality we put their eigenvalue equals
to zero, $E_0=0$. It is nothing but a choice of the reference
point for eigenvalues. Now we can construct a perturbation theory
for Riccati equation taking $\Psi_0$ and $y_0, V_0$ as zero
approximation, which characterizes the unperturbed problem. One
can write the original potential $V = m^2 x^2 + g x^4$ as a sum,
\begin{equation}
\label{VPT}
  V = V_0 + (V-V_0)\ \equiv\ V_0 + V_1\ ,
\end{equation}
thus, taking a deviation of the original potential from the
potential of the zero approximation as a perturbation. We always
can insert a formal parameter $\la$ in front of $V_1$ and develop
a perturbation theory in powers of $\la$,
\begin{equation}
\label{PT}
  E=\sum_{k=0}^{\infty} \la^k E_k \quad ,\quad y=\sum_{k=0}^{\infty}
  \la^k y_k \ ,
\end{equation}
putting $\la=1$ afterwards. Perhaps, it is worth emphasizing that
in spite of the fact that we study iteratively the equation
(\ref{RicAHO}), in general, this perturbation series has nothing
to do with a standard WKB expansion. By substituting (\ref{PT})
into (\ref{RicAHO}) we arrive at the equations which defines
iteratively the corrections
\begin{equation}
\label{PT-equation}
  {y}_k' -2 y_0 y_k = E_k - Q_k \ ,
\end{equation}
where
\begin{eqnarray*}
Q_1 &=& V_1 \ , \\
Q_k &=& -\sum_{i=1}^{k-1} y_i \cdot y_{k-i}\ ,\quad k=2,3,\ldots \ .
\end{eqnarray*}
It is interesting that the operator in the l.h.s. of
(\ref{PT-equation}) does not depend on $k$, while $Q_k$ in the
r.h.s. can be interpreted as a perturbation on the level $k$. The
solution of (\ref{PT-equation}) can be found explicitly and is
given by
\begin{eqnarray}
 E_k & = & \frac{\int_{-\infty}^{\infty} Q_k \Psi_0^2 \,
       dx}{\int_{-\infty}^{\infty}\Psi_0^2 \, dx}\ ,
\\[10pt]
 y_k & = & \Psi_0^{-2} \int_{-\infty}^{x} (E_k - Q_k) \Psi_0^2 \,
       dx' \ .
\end{eqnarray}
It is easy to demonstrate that if the first correction $y_1$ is bounded,
\begin{equation}
\label{convergence}
 |y_1| \leq \mbox{Const}\ ,
\end{equation}
it provides a sufficient condition for this perturbation theory
(\ref{PT}) to be convergent \cite{Turbiner:1984}. Note that this
condition is very rough and very likely can be strengthened.

({\bf ii}) The first two terms in the expansion of energy (\ref{PT})
in the above-described perturbation theory admit an interpretation
in the framework of the variational calculus \cite{Turbiner:1980}.
Let us assume that our variational trial function $\Psi_0(x)$ is
normalized to 1. We can calculate the potential $V_0$ where $\Psi_0(x)$
is the ground state eigenfunction and even put $E_0=0$
(see a discussion above). Formally, we construct the Hamiltonian
$H_0=p^2 + V_0$ for which $H_0 \Psi_0(x)=0$. The variational energy
is equal to
\begin{eqnarray}
\label{Evar}
  E_{var} & = & \int \psi_0 H \psi_0 = \underbrace{\int \psi_0 H_0 \,
  \psi_0}_{=E_0} +
  \underbrace{\int \psi_0 \underbrace{(H-H_0)}_{V-V_0} \psi_0}_{=E_1} \non
  \\
  & = & E_0 + E_1 (V_1=V-V_0) \geq E_{exact} \ .
\end{eqnarray}
Of course, $\Psi_0(x)$ could depend on free parameters. In this
case both $V_0$ and $V_1$ depend on parameters as well.
Minimization of $E_{var}$ with respect to the parameters can be
performed and the variational principle guarantees that $E_{var}$
gives upper bound to the ground state energy. This simple
interpretation (\ref{Evar}) reveals a fundamental difference
between perturbation theory and variational calculus. Variational
estimates can be obtained independently on the fact that the
perturbation theory associated with trial function $\Psi_0(x)$ is
convergent or divergent. However, it seems natural to remove this
difference by requiring a convergence of the perturbation series.
In this case by calculating the next terms $E_2,E_3,\ldots$ in
(\ref{PT}) one can estimate the accuracy of variational calculation
from one side and improve it iteratively from another side.
An immediate criteria how to choose $\Psi_0(x)$ in order to get
a convergent perturbation theory is to have the perturbation potential
$V_1$ to be subordinate with respect to the non-vanishing potential of zero
approximation $V_0$,
\begin{equation}
\label{PT-converge}
 \Big|\frac{V_1}{V_0} \Big| < 1 \qquad ,
 \qquad \mbox{for}\ |x| > R \quad .
\end{equation}
From this point of view any function (\ref{psi0}),(\ref{psi0+}) or, (\ref{psi0-})
taken as an entry leads to a convergent perturbation theory. An open question is
how to estimate the radius of convergency.

The requirement (\ref{PT-converge}) has a non-trivial physical implication:
in order to guarantee a convergence of perturbation theory a
domain where the wavefunction is exponentially small (classically-prohibited
domain) should be reproduced as precise as possible. The same time
a description of a domain where the wavefunction is of the order 1 is not
important. It contradicts to a straightforward physics intuition
and underlying idea of variational calculus which, in particular,
requires a precise description of the domain where the
wavefunction is of the order 1. Needless to say that namely the
latter domain gives a dominant contribution to the integrals which
define the energy in the variational calculations. Similar conclusion
was presented in \cite{Feynman:1987}.

{\bf \large Results.}

\begin{center}
    {\it \large Ground state}
\end{center}

\begin{center}
    {\bf $a=1$}
\end{center}

\noindent
This is the case of a single-well potential (anharmonic oscillator). The variational parameters 
in (\ref{psi0+}) are
\[
  D\ =\ 4.33441\ ,\ A\ =\ -9.23456\ ,\  \al \ =\ 2.74573\ .
\]
Ground state energy is
\[
  E_{var}\ (\equiv \ E_0+E_1)\ =\ 1.607541302594\ ,
\]
while the first correction to it is
\[
  \Delta E_{var}\ (\equiv \ E_2)\ =\ -1.2552 \times 10^{-10}\ .
\]
Eventually,
\begin{equation}
\label{a=1}
    \tilde E_{var} = E_{var} + \Delta E_{var}\ =\ 1.607541302469\ ,
\end{equation}
where all 13 digits are correct, since the next correction $E_3 \sim 10^{-14}$.
The rate of convergency seems extremely high, $\sim 10^{-4}$! The dependence the
energy on $\al$ is very weak: the results are almost unchanged if $\al=0$ (and (\ref{psi0+}) 
becomes (\ref{psi0})). It is the case for $a \geq 0$.

\begin{center}
    {\bf $a=-1$}
\end{center}

\noindent
This is already the case of a double-well potential. The variational parameters in 
(\ref{psi0+}) are
\[
  D\ =\ 4.059888\ ,\ A\ =\ -12.4816\ ,\  \al \ =\ 3.07041\ .
\]
In comparison with the case $a=1$ the parameters $D, \al$ are slightly changed unlike 
the parameter $A$ which decreased in $\sim 40 \%$.
Ground state energy is
\[
  E_{var}\ (\equiv \ E_0+E_1)\ =\ 1.029560832093\ ,
\]
while the first correction to it is
\[
  \Delta E_{var}\ (\equiv \ E_2)\ =\ -1.0382 \times 10^{-9}\ .
\]
Eventually,
\begin{equation}
\label{a=-1}
    \tilde E_{var} = E_{var} + \Delta E_{var}\ =\ 1.029560831054\ ,
\end{equation}
where all 13 digits are correct, since the next correction $E_3 \sim 10^{-13}$.
Similar to the case $a=1$ the rate of convergency seems extremely high, $\sim 10^{-4}$!

On Fig.1 one can see the behavior of logarithmic derivative $y_0$ vs $x\geq 0$. It is a 
very smooth function. The first correction $|y_1|$ (see Fig.2) has very interesting behavior: 
it is of the order of $10^{-4}$ at $x \lesssim 1$ (in the domain which give a dominant contribution 
to the energy integral (\ref{Evar})), then it starts to grow and reaches the maximum $\sim 0.006$ 
at $x \sim 3.9$ in the domain which gives negligibly small contribution to the energy integral, 
$y_0(x=3.9) \sim 15.2$ and $\Psi_0(x=3.9) \sim 10^{-9}$. In this case $\de \sim max|y_1|\sim 0.006$ 
(see p.1). Notably, $y_1 \propto 1/x^2$ at $x \gg 1$.
Similar behavior is demonstrated by the higher corrections $|y_n|, n=2,3$: they are very small at 
$x \lesssim 1$, its maximum is reached for $x > 1$ but its position is systematically reduced with $n$.

%%%%%%%%%%%%%% FIGURE 1
\begin{figure}
\begin{center}
  \includegraphics[width=1.8in,angle=0]{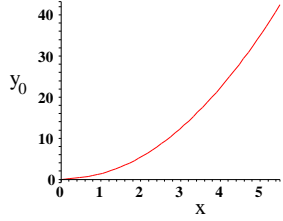}
    \caption{Logarithmic derivative $y_0 = \vphi_{int}'$ (see (\ref{phase_int})) as function 
    of $x$ for double-well potential (2) with $a=-1$ }
\end{center}
\end{figure}

%%%%%%%%%%%%%% FIGURE 2
\begin{figure}
\begin{center}
  \includegraphics[width=1.8in,angle=0]{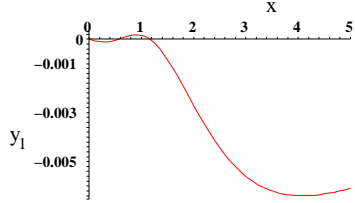}
    \caption{The first correction $y_1$ for $a=-1$ (cf. Fig.1)}
\end{center}
\end{figure}

\begin{center}
    {\bf $a=a_{crit}$}
\end{center}

\noindent
Let us consider the Hamiltonian (1). The ground state function is
symmetric w.r.t. $x \rar -x$. Hence, it has to have an extremum at
$x=0$. For any fixed $g$ there exists
a value $m^2_{crit} < 0$ such that for $m^2 > m^2_{crit}$ this extremum
is a maximum, otherwise a minimum. It is easy to find out that the
critical point $m^2_{crit}$ corresponds to the vanishing ground
state energy, $E=0$ \footnote{In classical case, it corresponds to 
the particle stopping on the top of the barrier.}.
Using the function (\ref{psi0+}) it was
calculated the critical value $m^2_{crit}=-2.2195970861$ for $g=1$.
For the Hamiltonian (\ref{AHOa})
\[
 E (a_{crit}=-3.523390749)\ =\ 0\ .
\]
It is quite interesting from physical point of view that for a family
of double-well potentials with fixed $g$ there exists a domain $0>
m^2 > (m^2)_{crit}$ where the ground-state eigenfunction has the
maximum at the origin, which corresponds to the position of the
unstable equilibrium similar to what takes place for the
single-well case. It implies that the particle in such a
potential with the ground state energy above the barrier, $E>0$,
somehow does not feel the existence of two minima. In this domain
WKB consideration at $x \sim 0$ is not valid.

\begin{center}
    {\bf $a=-20$}
\end{center}

\noindent
This is the case of a double-well potential. The variational parameters 
in (\ref{psi0+}) are
\[
  D\ =\ 6.765663\ ,\ A\ =\ -286.6456\ ,\  \al \ =\ 49.6136\ .
\]
Ground state energy is
\[
  E_{var}\ (\equiv \ E_0+E_1)\ =\ -43.7793127\ ,
\]
while the first correction to it is
\[
  \Delta E_{var}\ (\equiv \ E_2)\ =\ -3.81 \times 10^{-6}\ .
\]
Eventually,
\begin{equation}
\label{a=20}
    \tilde E_{var} = E_{var} + \Delta E_{var}\ =\ -43.7793165\ ,
\end{equation}
where all 9 digits are correct, since the next correction $E_3 \sim 10^{-8}$.
Similar to the cases $a=\pm 1$ the rate of convergency seems still high, 
$\sim 10^{-2}$!

\begin{center}
    {\it \large First excited state}
\end{center}

\begin{center}
    {\bf $a=-20$}
\end{center}

\noindent
This is the case of a double-well potential. The variational parameters in 
(\ref{psi0-}) are
\[
  D\ =\ 5.584376\ ,\ A\ =\ -246.64375\ ,\  \al \ =\ 38.82768\ .
\]
The parameters seems quite close to those for $a=20$ for the ground state.
Ground state energy is
\[
  E_{var}\ (\equiv \ E_0+E_1)\ =\ -43.77931637\ ,
\]
while the first correction to it is
\[
  \De E_{var}\ (\equiv \ E_2)\ =\ -9.3618 \times 10^{-8}\ .
\]
Eventually,
\begin{equation}
\label{a=20-1st}
    \tilde E_{var} = E_{var} + \Delta E_{var}\ =\ -43.77931646\ ,
\end{equation}
where all 10 digits are correct, since the next correction $E_3 \sim 10^{-10}$.
Similar to the cases $a=20$ for the ground state the rate of convergency seems 
still high, $\sim 10^{-2}$!

\centerline{\large \it Energy Gap}

By definition the energy gap is
\[
    \De E \ =\ E_{first\ excited\ state}\ -\ E_{ground\ state}\ .
\]
For the double-well potential (2) at $a \rar -\infty$ it can be calculated and 
in one-instanton approximation it reads \cite{LL-QM:1977, Zinn:2005, Shuryak:1994}
\begin{equation}
\label{expansion}
    \De E \ =\ \frac{2^{11/4}}{\sqrt{\pi}} |a|^{5/4} e^{-
                 \frac{\sqrt{2}|a|^{3/2}}{6}}\bigg(1 -
                 \frac{71}{12} \frac{1}{\sqrt{2}|a|^{3/2}}
                 - \frac{6299}{288} \frac{1}{2|a|^{3}}
                 - \frac{2691107}{10368} \frac{1}{2\sqrt{2}|a|^{9/2}}
                 - \frac{2125346615}{497664} \frac{1}{4|a|^{6}}
                 \ldots \bigg)
\end{equation}
The first term in the expansion is a common knowledge.
All other coefficients in the expansion are due to J Zinn-Justin, 1981-2005
(see \cite{Zinn:2005} and references therein), they were obtained using the
so called {\it exact Bohr-Sommerfeld quantization condition}.
The coefficient 71/12 was independently calculated (and confirmed) in
two-loop instanton calculation by E Shuryak (see \cite{Shuryak:1994} and
references therein). It seems highly desirable to perform three-loop calculation
to check the next coefficient in the expansion. It is an asymptotic expansion 
\cite{Zinn:2005}.

It is interesting to compare the energy gap $\De E$ at $a=-20$ calculated in the 
(convergent) perturbation theory with (\ref{psi0+}) and (\ref{psi0-}) as zero 
approximations and with use of the (asymptotic) expansion (\ref{expansion}). Subsequent 
expressions show how $\De E$ evolves from pure variational results to ones with the first 
corrections $E_2$ taken into account and then to ones with the second corrections 
$E_3$ involved
\[
    \De E_{var} \ =\ 1.03282 \times 10^{-7}\ ,
\]
\[
    \De E^{(1)}_{var} \ =\ 1.06529 \times 10^{-7}\ ,
\]
\begin{equation}
\label{DelE-PT}
    \De E^{(2)}_{var} \ =\ 1.06525 \times 10^{-7}\ .
\end{equation}
In the last expression $\De E^{(2)}_{var}$ all six significant digits are correct.
Now how the energy gap $\De E$ looks like as function of a number of corrections to 
one-instanton result included (see (\ref{expansion})) 
\footnote{three and more instanton contributions are negligibly small for $a=-20$}:
\[
 one-instanton = 1.12154\times 10^{-7}\qquad (5.3\%\ deviation)
\]
\[
 one-instanton + 1st\ correction = 1.06908\times 10^{-7}\ (0.36\%\ deviation)
\]
\[
 one-instanton + 1st\ and\ 2nd\ corrections = 1.06754\times 10^{-7}\ (0.22\%\ deviation)
\]
\begin{equation}
\label{DelE-Inst}
 one-instanton + four\ corrections = 1.06738 \times 10^{-7}\ (0.20\%\ deviation)
\end{equation}
The numbers in brackets are relative deviations from (\ref{DelE-PT}). The fact that 
a deviation stays almost the same after adding 2nd, 3rd and 4th corrections likely
indicates the maximal accuracy based on a use of asymptotic expansion is reached and in 
any moment the result can blow up. A comparison of two values (\ref{DelE-PT}) and 
(\ref{DelE-Inst}) shows that they do not agree in the 4th digit.

In a conclusion I have to say that a similar consideration based on interpolation of 
phase between small and large distances was done for sextic oscillator and the Zeeman
effect on hydrogen. In both cases the exceptionally high accuracies were obtained.

{\bf Acknowledgments}\\
Author thanks E.~Shuryak, A.I.~Vainshtein, J.C.L.~Vieyra for valuable discussions.
The work is supported in part by grants: CONACyT {\it 58942-F}, DGAPA {\it IN115709-3}.

\begingroup\raggedright
\endgroup


\begin{thebibliography}{99}

\bibitem{Polyakov:1977}
        A.M.~Polyakov, {\it Nucl.Phys, \bf B122,} 429 (1977)

\bibitem{Coleman:1979}
        S.R.~Coleman,
        %``The uses of instantons,''
        {\it Subnucl.\ Ser.\ \bf 15}, 805 (1979)

\bibitem{Vainshtein:1981}
%        A.I.~Vainshtein, V.I.~Zakharov, V.A.~Novikov and M.A.~Shifman,
%        %``ABC of instantons,''
%        {\it Sov.\ Phys.\ Usp.\ \bf 25}, 195 (1982)
%        [Usp.\ Fiz.\ Nauk {\bf 136}, 553 (1982)],
        M.A.~Shifman, ``Instantons in Gauge Theories" (World Scientific, 
        Singapore, 1994)

%\bibitem{Zinn-Justin:2002}
%        J. Zinn-Justin,
%   	    ``Quantum Field Theory and Critical Phenomena",
%         4th ed. (Clarendon Press, Oxford, 2002), pp. 1024

\bibitem{BW}
         C.M.~Bender, T.T.~Wu,
%         `Anharmonic Oscillator',
         {\it Phys. Rev. \bf 184}, 1231 (1969);
%         `Anharmonic Oscillator.II',
         {\it Phys. Rev.  \bf D 7} , 1620 (1973)

\bibitem{Turbiner:1980}
        A.V.~Turbiner,
%        `On Perturbation Theory and Variational Methods in Quantum
%           Mechanics', \\
           {\it Soviet Phys. -- ZhETF \bf 79}, 1719-1745 (1980);\\
           {\it JETP \bf 52}, 868-876 (1980)
           (English Translation)

\bibitem{Eremenko:2008}
         A.~Eremenko, A.~Gabrielov, B.~Shapiro,
%         `Zeros of Eigenfunctions of some Anharmonic Oscillators',\\
         {\it Ann. Inst. Fourier, Grenoble \bf 58}, 603-624 (2008)

\bibitem{Turbiner:1984}
        A.V.~Turbiner,
%        `The Problem of Spectra in Quantum Mechanics and the
%        `Non-Linearization' \\ Procedure', \\
        {\it Usp. Fiz. Nauk. \bf 144}, 35-78 (1984),\\
        {\it Sov. Phys. - Uspekhi \bf 27}, 668-694 (1984)
        (English Translation)

\bibitem{Turbiner:2005}
        A.V.~Turbiner,
%        `Anharmonic oscillator and double-well potential:
%            approximating eigenfunctions'
%         Preprint ICN-UNAM 05-03, pp.14 (June 2005)\\
%           {\it (invited contribution to the memory of Felix~A.~Berezin)}\\
%                 (math-ph/0506033)\\
           {\it Letters in Mathematical Physics \bf 74}, 169-180 (2005)

\bibitem{LL-QM:1977}
        L.D.~Landau and E.M.~Lifshitz,
        {\it Quantum Mechanics},
        Pergamon Press (Oxford - New York - Toronto -
        Sydney - Paris - Frankfurt), 1977

\bibitem{Price}
          P.J.~Price, {\it Proc. Phys. Soc. London \bf 67}, 383 (1954)

\bibitem{Turbiner:1979}
        A.V.~Turbiner,
%        `A New Approach to Finding Levels of Energy of Bound States\\
%           in Quantum Mechanics:  Convergent Perturbation Theory',\\
           {\it Soviet Phys. -- Pisma ZhETF \bf 30}, 379-383 (1979).\\
           {\it JETP Lett. \bf 30}, 352-355 (1979)
           (English Translation)

\bibitem{Feynman:1987}
        R.P.~Feynman,
      `Difficulties In Applying The Variational Principle To Quantum 
       Field Theories',
       in PROCEEDINGS of Int. Workshop on Variational Calculus in 
       Quantum Field Theory,
        Wangerooge, West Germany, Sept. 1-4, 1987\\
       (World Scientific, Singapore, 1987) pp. 28-40

\bibitem{Zinn:2005}
         J.~Zinn-Justin and U.D.~Jentschura,
%         `Multi-Instantons and Exact Results II: Specific Cases, Higher-Order Effects,
%         and Numerical Calculations',\\
         {\it Annals Phys. \bf 313}, 269-325 (2004);
         quant-ph/0501137 (updated, January 2005)

\bibitem{Shuryak:1994}
         C.E.~W\"ohler, E.~Shuryak,
%         `Two-loop correction to the instanton density for the double well
%          potential',\\
          {\it Phys. Lett. \bf B 333} (1994) 467--470

\end{thebibliography}
\end{document}